%% file: sync_icassp.tex
\documentclass{article}
\usepackage{spconf}
\usepackage{amsthm}
\input{SupportDocuments/YFPreamble.tex}

\usepackage{winsnotation}

\title{GENERALIZED DETERMINISTIC-RANDOM TRADEOFF IN INTEGRATED SENSING AND COMMUNICATIONS: THE SENSING-OPTIMAL OPERATING POINT}
\name{Yifeng~Xiong$^{\dag}$ \qquad Fan Liu$^{\ddag}$\qquad Marco Lops$^{\sharp}$
\thanks{Corresponding author: Fan Liu.}
\thanks{E-mail:liuf6@sustech.edu.cn}
}

\address{
$^{\dag}$ SICE, Beijing University of Posts and Telecommunications, China \\
$^{\ddag}$Department of EEE, Southern University of Science and Technology, China \\
$^{\sharp}$ DIETI, University of Naples Federico II, Italy \& CNIT, Italy
}

\selectfont
\begin{document}
\maketitle

\begin{abstract}
Integrated sensing and communications (ISAC) has been recognized as a key component in the envisioned 6G communication systems. Understanding the fundamental performance tradeoff between sensing and communication functionalities is essential for designing practical cost-efficient ISAC systems. In this paper, we aim for augmenting the current understanding of the deterministic-random tradeoff (DRT) between sensing and communication, by analyzing the sensing-optimal operating point of the fundamental capacity-distortion region. We show that the DRT exists for generic sensing performance metrics that are in general not convex/concave in the ISAC waveform. Especially, we elaborate on a representative non-convex performance metric, namely the detection probability for target detection tasks.
\end{abstract}


\section{Introduction}
\Ac{isac} has been recognized as one of the six typical usage scenarios of the \ac{6g} wireless system in the recently released \ac{itu} \ac{6g} vision \cite{vision_6g}, due to its potential of offering a new channel linking the physical world to the digital world via the mutual benefit between sensing and communication functionalities. A large body of literature has been devoted to the design of \ac{isac} systems \cite{9737357,9093221,liu2021CRB,9529026,isac_11ad1,isac_11ad2}. By contrast, many theoretical research challenges remain, among which the most fundamental one is the characterization of the \ac{snc} performance tradeoff, which originates from the fact that \ac{snc} functionalities are realized relying on a unified waveform and the shared use of wireless resources.

In its most general form, the \ac{snc} tradeoff is characterized by the Pareto front determined by the following generic optimization problem
\begin{equation}\label{snc_tradeoff}
\begin{aligned}
\max_{p_{\RM{X}}(\M{X})}&~~ \frac{1}{T}I(\RM{Y}_{\rm c};\RM{X}|\RM{H}_{\rm c}),\\
{\rm s.t.}&~~\mathbb{E}\{\rv{e}(\RM{X})\}\leq E,\\
&~~\mathbb{E}\{\rv{c}(\RM{X})\} \leq C,
\end{aligned}
\end{equation}
where $\rv{e}(\RM{X})$ represents a certain metric of sensing performance, e.g., mean squared error (MSE), while the objective function is the achievable communication rate under the channel $\RM{H}_{\rm c}$, with $\RM{X}\in\mathbb{C}^{M\times T}$ and $\RM{Y}_{\rm c}\in\mathbb{C}^{N_{\rm c}\times T}$ denote the transmitted \ac{isac} signal (which is random but known at the sensing receiver), and the received signal at the communication receiver, respectively. The function $\rv{c}(\RM{X})$ denotes the resource cost, e.g. power, bandwidth, etc. As an important example, in \cite{tit_cd,isit_cd1,isit_cd2}, it has been shown that any expected sensing distortion constraint taking the following form
$$
\mathbb{E}\{\rv{d}(\RV{\eta},\hat{\RV{\eta}}(\RM{X},\RM{Y}_{\rm s}))\}\leq D,
$$
can be equivalently expressed in the form of the expected sensing metric constraint $\mathbb{E}\{\rv{e}(\RM{X})\}\leq E$, where $\RV{\eta}$ represents the sensing parameter, and $\hat{\RV{\eta}}(\RM{X},\RM{Y}_{\rm s})$ denotes its estimate based on the sensing received signal $\RM{Y}_{\rm s}$.

Alternatively, $\rv{e}(\RM{X})$ can also be any reasonable sensing performance metric that does not correspond explicitly to a distortion metric, for example, the \ac{crb} for estimation problems. It has been revealed in \cite{tit_crb_rate} that the \ac{crb}-rate tradeoff is two-fold, consisting of the \ac{st} determined by resource allocation and the \ac{drt} related to the signalling strategies. To elaborate, \ac{drt} refers to the phenomenon that communication favors random signals, while sensing favors deterministic signals. In particular, the \ac{crb}-rate \ac{drt} follows from the convexity of \ac{crb}: the sensing-optimal operating point corresponds to a deterministic sample covariance matrix $\RM{R}_{\RM{X}}=T^{-1}\RM{X}\RM{X}^{\rm H}$, according to the Jensen's inequality, leading to a communication \ac{dof} loss.

A natural follow-up question is that, whether \ac{drt} exists for other sensing performance metrics. In \cite{est_rate}, the authors have shown that the negative mutual information $-I(\RM{Y}_{\rm s};\RM{\eta}|\RM{X})$ is convex in $\RM{R}_{\RM{X}}$, and hence admits a \ac{drt}. Furthermore, by viewing the sensing process as non-cooperative joint source-channel coding (with the environment to be sensed being the encoder) \cite{drt_mi}, the mutual information is shown to set a lower bound for any distortion metric, suggesting that \ac{drt} may exist whenever the bound is achievable. Unfortunately, the bound is seldom achievable, since the ``coding capability'' of the environment is highly limited.

In this paper, we tackle the problem by directly analyzing the generic problem \eqref{snc_tradeoff}. We show that at the sensing-optimal operating point $P_{\rm SC}$, the communication performance loss is essentially a consequence of the \textit{solution sparsity of linear functional programming}, and hence the existence of \ac{drt} does not depend on the convexity of sensing performance metrics. As an important example, we conduct a case study on the target detection problem using \ac{cfar} detectors. In this scenario, the sensing performance metric is the detection probability, which does not have a constant convexity in $\RM{X}$, while the \ac{drt} persists.

\section{General Result}
\begin{figure}[t]
    \centering
    \begin{overpic}[width=.4\textwidth]{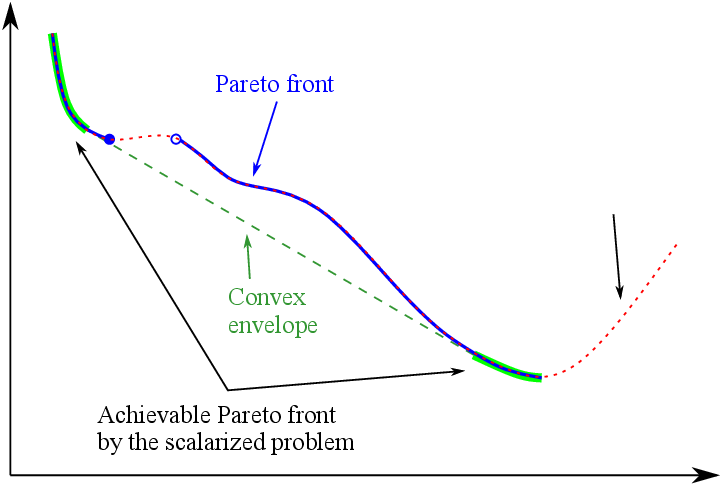}
    \put(-8,64){\scriptsize $e(\M{X})$}
    \put(102,0){\scriptsize $c(\M{X})$}
    \put(74,45){\scriptsize $\begin{aligned}\min_{\M{X}}&~e(\M{X}),\\{\rm s.t.}&~c(\M{X})=C\end{aligned}$}
    \end{overpic}
    \caption{\small The optimal solutions to the equality-constrained problem \eqref{equality_sensing_cost}, the Pareto front obtained by solving \eqref{pareto_sensing_cost} (or equivalently \eqref{determinisitic_sensing_cost}),  and the part achievable by the scalarized problem \eqref{scalarized_sensing_cost}.}
    \label{fig:fronts_illustration}
\end{figure}

To characterize the sensing-optimal operating point $P_{\rm SC}$, we consider the optimal sensing performance given by the following problem
\begin{equation}\label{general_sensing_cost}
\begin{aligned}
\min_{p_{\RM{X}}(\M{X})}&~~\mathbb{E}\{\rv{e}(\RM{X})\},\\
{\rm s.t.}&~~\mathbb{E}\{\rv{c}(\RM{X})\}\leq C.
\end{aligned}
\end{equation}
We are interested in the structure of the sensing-optimal $p_{\RM{X}}(\M{X})$ obtained by solving \eqref{general_sensing_cost}, which enforces constraints on the achievable communication rate. Before delving into details, we would like to provide some intuitions by contrasting \eqref{general_sensing_cost} with the associated deterministic optimization problem given by
\begin{equation}\label{determinisitic_sensing_cost}
\begin{aligned}
\min_{\M{X}}&~~e(\M{X}),\\
{\rm s.t.}&~~c(\M{X})\leq C.
\end{aligned}
\end{equation}
The problem \eqref{determinisitic_sensing_cost} can be seen to be equivalent to the following multi-objective optimization problem
\begin{equation}\label{pareto_sensing_cost}
\min_{\M{X}}~~\left(e(\M{X}),c(\M{X})\right),
\end{equation}
which is in turn related to the following scalarized problem
\begin{equation}\label{scalarized_sensing_cost}
\min_{\M{X}}~~e(\M{X}) + \lambda c(\M{X}),
\end{equation}
for some $\lambda > 0$, as well as the equality-constrained problem
\begin{equation}\label{equality_sensing_cost}
\begin{aligned}
\min_{\M{X}}&~~e(\M{X}),\\
{\rm s.t.}&~~c(\M{X})= C.
\end{aligned}
\end{equation}
Under different choices of $C$ and $\lambda$, the optimal achievable $e(\M{X})$ and $c(\M{X})$ for these problems are portrayed in Fig.~\ref{fig:fronts_illustration}. 

For better clarity, let us refer to an instance of $\RM{X}$ as a \emph{pure signalling strategy}. The Pareto front shown in Fig.~\ref{fig:fronts_illustration} depicts the optimal sensing performance that is achievable by all pure signalling strategies under resource constraints. Using this terminology, the distribution $p_{\RM{X}}(\M{X})$ may be interpreted as a time-sharing scheme over pure signalling strategies, and thereby produces a \emph{mixed signalling strategy}. Intuitively, mixed signalling strategies should be capable of achieving the convex envelope of the region achieved by the pure strategies, as portrayed in Fig.~\ref{fig:fronts_illustration}. Next, we present a formal version of the aforementioned intuition, by constructing explicitly the sensing-optimal signalling strategy $p_{\RM{X}}(\M{X})$.
\begin{proposition}[Sensing-optimal $p_{\RM{X}}(\M{X})$]\label{prop:optimal_px}
Given an expected resource budget $C$, the sensing-optimal distribution $p_{\RM{X}}(\M{X})$ can be decomposed as
\begin{equation}
p_{\RM{X}}(\M{X}) = \int p_{\RM{X}|\rv{c}(\RM{X})}(\M{X}|c(\M{X})) p_{\rv{c}(\RM{X})}(c(\M{X})){\rm d} c(\M{X}),
\end{equation}
where the conditional distribution $p_{\RM{X}|\rv{c}(\RM{X})}(\M{X}|c(\M{X}))$ can be an arbitrary probability distribution satisfying
\begin{equation}\label{optimal_conditional}
\begin{aligned}
&{\rm supp}\left(p_{\RM{X}|\rv{c}(\RM{X})}(\M{X}|c(\M{X})=\xi)\right) \\
&\hspace{3mm}= \left\{\M{X}\Big|e(\M{X})=\min\limits_{\M{X}}~e(\M{X}),~{\rm s.t.}~c(\M{X})=\xi\right\}\\
&\hspace{3mm}:=\Set{S}_{\rm opt}(\xi).
\end{aligned}
\end{equation}
The distribution $p_{\rv{c}(\RM{X})}(c(\M{X}))$ takes the following form
\begin{equation}\label{optimal_pcx}
p_{\rv{c}(\RM{X})}(c(\M{X})) \!=\! \left\{
\begin{array}{ll}
\delta\left(c(\M{X})\!-\!C\right), & \hbox{$C\in\widetilde{\Set{G}}$;}\\
\delta\left(c(\M{X})\!-\!\max\limits_{\xi\in\widetilde{\Set{G}}}~\xi\right), & \hbox{$C\!\geq\! \max\limits_{\xi\in\widetilde{\Set{G}}}~\xi$;}\\
q(c(\M{X})), & \hbox{$C\!\notin\!\widetilde{\Set{G}}$, $C\!<\!\max\limits_{\xi\in\widetilde{\Set{G}}}~\xi$,}
\end{array}
\right.
\end{equation}
where $q(c(\M{X}))$ can be an arbitrary probability distribution satisfying the following conditions
\begin{subequations}
\begin{align}
&{\rm supp}(q(c(\M{X})))\subseteq \Set{S}[\tilde{g},C], \label{qcx_condition_supp}\\
&\int c(\M{X}) q(c(\M{X})){\rm d}c(\M{X}) = C, \label{qcx_condition_mean}\\
&\int q(c(\M{X})) {\rm d}c(\M{X}) =1,~q(c(\M{X}))\geq 0~\forall c(\M{X})\geq 0. \label{qcx_condition_prob}
\end{align}
\end{subequations}
The set $\Set{S}[\tilde{g},C]$ is uniquely determined by $\tilde{g}(\cdot)$ and $C$ via the following conditions
\begin{subequations}\label{tangents}
\begin{align}
&\lambda\geq 0,~\mu\geq 0,\\
&\Set{S}[\tilde{g},C] = \{\xi|-\tilde{g}(\xi) = \lambda+\mu \xi\},~|\tilde{S}[\tilde{g},C]|>0,\\
&\forall \xi\geq 0,~-\tilde{g}(\xi)\leq \lambda+\mu\xi,\\
&\min_{\xi\in\Set{S}[\tilde{g},C]}~\xi \leq C\leq \max_{\xi\in\Set{S}[\tilde{g},C]}~\xi.
\end{align}
\end{subequations}
$\tilde{g}(\cdot)$ denotes the subset of the Pareto front for problem \eqref{pareto_sensing_cost} that is achievable by the scalarized problem \eqref{scalarized_sensing_cost}, namely
\begin{equation}
\tilde{g}(c(\M{X})) = \left\{
\begin{array}{ll}
e(\M{X}), & \hbox{$c(\M{X})\in\widetilde{\Set{G}}$;} \\
{\rm undefined}, & \hbox{otherwise,}
\end{array}
\right.
\end{equation}
while $\widetilde{\Set{G}}$ denotes the domain of the function $\tilde{g}(\cdot)$, given by
\begin{equation}\label{setg}
\begin{aligned}
&\widetilde{\Set{G}}=\Big\{c(\M{X})\Big|e(\M{X}) + \lambda c(\M{X})=\\
&\hspace{3mm}\mathop{\rm min}\limits_{\M{X}}\left[e(\M{X}) + \lambda c(\M{X})\right],~\lambda\geq 0\Big\}.
\end{aligned}
\end{equation}
\begin{proof}
Please refer to Appendix \ref{proof_optimal_px}.
\end{proof}
\end{proposition}

\begin{remark}
Observe that the original problem \eqref{determinisitic_sensing_cost} (or \eqref{pareto_sensing_cost}) is a linear functional programming with respect to $p_{\RM{X}}(\M{X})$. Geometrically speaking, the conditions \eqref{tangents} determine the tangent points of the Pareto front for problem \eqref{pareto_sensing_cost} to its convex envelope obtained by the mixed signalling strategies. Consequently, only the points shared by the Pareto front and its convex envelope could serve as candidates for sensing-optimal distributions. This implies that besides the convexity of sensing performance metrics, the sparsity of the expected sensing cost minimization problem is also a source of the \ac{drt}.
\end{remark}

Next, we consider the achievable communication rate at point $P_{\rm SC}$. First note that
\begin{equation}
\begin{aligned}
I(\RM{Y}_{\rm c};\RM{X}|\RM{H}_{\rm c}) &= I(\RM{Y}_{\rm c};\rv{c}(\RM{X})|\RM{H}_{\rm c}) + I(\RM{Y}_{\rm c};\RM{X}|\rv{c}(\RM{X}),\RM{H}_{\rm c}) \\
&=I(\RM{Y}_{\rm c};\rv{c}(\RM{X})|\RM{H}_{\rm c}) \\
&\hspace{3mm}+ \int I(\RM{Y}_{\rm c};\RM{X}|\rv{c}(\RM{X})=\xi,\RM{H}_{\rm c}) p_{c(\RM{X})}(\xi){\rm d}\xi.
\end{aligned}
\end{equation}
The maximum achievable rate at $P_{\rm SC}$ can thus be written as
\begin{subequations}
\begin{align}
R_{\rm SC} &\!=\! \frac{1}{T} \max_{p_{\RM{X}}(\M{X})}~I(\RM{Y}_{\rm c};\RM{X}|\RM{H}_{\rm c}),~~{\rm s.t.}~\eqref{optimal_conditional},\eqref{optimal_pcx}\\
&\!=\!\frac{1}{T}\max_{p_{\RM{X}}(\M{X})}~\Big\{\int I(\RM{Y}_{\rm c};\RM{X}|\rv{c}(\RM{X})\!=\!\xi,\RM{H}_{\rm c}) p_{c(\RM{X})}(\xi){\rm d}\xi \nonumber \\
&\hspace{3mm}\!+\! I(\RM{Y}_{\rm c};\rv{c}(\RM{X})|\RM{H}_{\rm c})\Big\},~~{\rm s.t.}~\eqref{optimal_conditional},\eqref{optimal_pcx},\label{opt_problem_rate}
\end{align}
\end{subequations}
from which it is clear that $R_{\rm SC}<\frac{1}{T}\max_{p_{\RM{X}}(\M{X})}I(\RM{Y}_{\rm c};\RM{X}|\RM{H}_{\rm c})$, since $R_{\rm SC}$ corresponds to a mutual information maximization problem with additional constraints. This implies the \ac{drt}: sensing optimality leads to communication rate loss.

To obtain more explicit expressions for $R_{\rm SC}$, according to Proposition \ref{prop:optimal_px}, when $C\in\widetilde{\Set{G}}$, we have
$$
R_{\rm SC} = \frac{1}{T} \max_{p_{\RM{X}|\rv{c}(\RM{X})}(\M{X}|c(\M{X})=C)}~I(\RM{Y}_{\rm c};\RM{X}|\rv{c}(\RM{X})=C,\RM{H}_{\rm c}).
$$
When $C\geq\max_{\xi \in\widetilde{\Set{G}}}~\xi$, we have
$$
\begin{aligned}
R_{\rm SC} &= \frac{1}{T} \max_{p_{\RM{X}|\rv{c}(\RM{X})}(\M{X}|c(\M{X})=\max_{\xi\in\widetilde{\Set{G}}}\xi)}\\
&\hspace{3mm}I\left(\RM{Y}_{\rm c};\RM{X}|\rv{c}(\RM{X})=\max_{\xi\in\widetilde{\Set{G}}}~\xi,\RM{H}_{\rm c}\right).
\end{aligned}
$$
In the case of $C\notin \widetilde{\Set{G}},~C<\max_{\xi\in\widetilde{\Set{G}}}~\xi$, in general, one have to solve the optimization problem \eqref{opt_problem_rate}. Fortunately, when $|\Set{S}[\tilde{g},C]|=2$, $p_{\rv{c}(\RM{X})}(c(\M{X}))$ can be uniquely determined using the conditions \eqref{qcx_condition_supp}-\eqref{qcx_condition_prob}. Specifically, we have
\begin{equation}
\begin{aligned}
p_{\rv{c}(\RM{X})}(c(\M{X})) &= \frac{\xi_2-C}{\xi_2-\xi_1}\delta\left(c(\M{X})-\xi_1\right) \\
&\hspace{3mm}+\frac{C-\xi_1}{\xi_2-\xi_1}\delta\left(c(\M{X})-\xi_2\right),
\end{aligned}
\end{equation}
where $\xi_1$ and $\xi_2$ satisfy $\Set{S}[\tilde{g},C]=\{\xi_1,\xi_2\}$ and $\xi_1<\xi_2$. Upon denoting $p_1=\frac{\xi_2-C}{\xi_2-\xi_1}$ and $p_2=\frac{C-\xi_1}{\xi_2-\xi_1}$, the optimization problem \eqref{opt_problem_rate} is simplified to
\begin{equation}
\begin{aligned}
R_{\rm SC}=\frac{1}{T}\max_{p_{\RM{X}|\rv{c}(\RM{X})}(\M{X}|c(\M{X}))}~\Big\{&I(\RM{Y}_{\rm c};\rv{c}(\RM{X})|\RM{H}_{\rm c}) + \\
&\sum_{i=1}^2 p_i I(\RM{Y}_{\rm c};\RM{X}|\rv{c}(\RM{X})=\xi_i,\RM{H}_{\rm c}) \Big\}.
\end{aligned}
\end{equation}

In fact, $|\Set{S}[\tilde{g},C]|=2$ holds for many practical Pareto fronts $\tilde{g}$. An important example will be presented in Sec. \ref{sec:detection}.

\section{Case Study: Binary Detection in MIMO Radar with Colocated Transmit and Receive Antennas}\label{sec:detection}
\subsection{Problem Formulation and Sensing Metrics}
In this section, we consider the target presence detection problem in \ac{mimo} Radar with colocated transmit and receive antennas. We assume that all the receive antennas see the target under the same aspect angle, which results in a unique, possibly random, complex scattering coefficient from the target. As a consequence, the model reads
\begin{equation}
\RM{Y}_{\rm s}=\rv{A}e^{j\rv{\theta}}\M{H}_{\rm s} \RM{X}+\RM{W} \; , \qquad \rv{\theta} \sim {\cal U}(-\pi, \pi)
\end{equation}
Notice that the matrix $\M{H}_{\rm s}$ is typically a Kronecker product and may contain target parameters of interest. In the presence of independent white noise, the Likelihood Ratio is
\[
\begin{aligned}
L_{\RM{Y}_{\rm s}}(\M{Y}_{\rm s}) &\propto \mathbb{E}_{\rv{A}}\Big[e^{-\frac{\rv{A}^2}{N_0}\sum_{i=1}^T\| \M{H}_{\rm s} \V{x}(i)\|^2}\\
&\hspace{3mm}\times I_0 \Big(\frac{2\rv{A}}{N_0}\Big|\sum_{i=1}^T\V{y}_{\rm s}^{\rm H}(i)\M{H}_{\rm s} \V{x}(i)\Big| \Big)
\Big]
\end{aligned}
\]
where $I_0(\cdot)$ is the modified Bessel function of order 0. This expression simplifies for Rayleigh fading, but in any case, whatever the density of $\rv{A}$, the optimum detector to ascertain the presence of a target reads:
\begin{equation}\label{cfar_detector}
\rv{Z}=\left| \sum_{i=1}^T \RV{y}^{\rm H}_{\rm s}(i)\M{H}_{\rm s} \RV{x}(i)\right|^2 \mathop{\gtrless}_{\mathcal{H}_0}^{\mathcal{H}_1}  \lambda.
\end{equation}
Assume now that $\rv{A}$ is Rayleigh, whereby $\rv{A}e^{j \rv{\theta}} \sim \mathcal{CN}(0, \overline{\rv{A}^2})$.
Notice that the test statistic is exponential under the two alternatives. In particular, since
$$
\begin{aligned}
\mathbb{E}\left[ \rv{Z}|\mathcal{H}_1\right]=&\overline{\rv{A}^2}T^2 {\rm Tr}^2\left( \M{H}_{\rm s} \RM{R}_{\RM{X}}\M{H}_{\rm s}^{\rm H}\right)+
T {\rm Tr}\left( \M{H}_{\rm s} \RM{R}_{\RM{X}}\M{H}_{\rm s}^{\rm H}\right)\\
\mathbb{E}\left[ \rv{Z}|\mathcal{H}_0\right]=& TN_0 {\rm Tr}\left( \M{H}_{\rm s} \RM{R}_{\RM{X}}\M{H}_{\rm s}^{\rm H}\right)
\end{aligned}
$$
with $\RM{R}_{\RM{X}}=\frac{1}{T}\sum_{i=1}^T \RV{x}(i) \RV{x}^{\rm H}(i)$ the signal sample covariance matrix, the \ac{snr} is thus written in the form
\[
{\rm SNR}=\frac{\overline{\rv{A}^2}T}{N_0}{\rm Tr}\left( \M{H}_{\rm s} \RM{R}_{\RM{X}}\M{H}_{\rm s}^{\rm H}\right)=\rv{\rho}
\]
We thus have
$$
\begin{aligned}
\rv{P}_{\rm FA}&=e^{\!-\!\frac{\lambda}{TN_0 {\rm Tr}\left( \M{H}_{\rm s} \RM{R}_{\RM{X}}\M{H}_{\rm s}^{\rm H}\right)} },\\
\rv{P}_{\rm d}&= e^{ \!-\!\frac{\lambda}{\overline{\rv{A}^2}T^2 {\rm Tr}^2\left( \M{H}_{\rm s} \RM{R}_{\RM{X}}\M{H}_{\rm s}^{\rm H}\right)\!+\!TN_0 {\rm Tr}\left( \M{H}_{\rm s} \RM{R}_{\RM{X}}\M{H}_{\rm s}^{\rm H}\right)}}
\end{aligned}
$$
yielding the well-known relationship
\[
\rv{P}_{\rm d}=\rv{P}_{\rm FA}^{\frac{1}{1+\rv{\rho}}}=\rv{P}_{\rm FA}^{\frac{1}{1+\frac{\overline{\rv{A}^2}T}{N_0}{\rm Tr}\left( \M{H}_{\rm s} \RM{R}_{\RM{X}}\M{H}_{\rm s}^{\rm H}\right)}}.
\]
For the \ac{cfar} detectors taking the form of \eqref{cfar_detector}, the false alarm rate $\rv{P}_{\rm FA}$ is a constant input parameter, and hence the detection probability $\rv{P}_{\rm d}$ is only a function of the \ac{snr} $\rv{\rho}$. In light of this, we shall denote the detection probability as $\rv{P}_{\rm d}(\rv{\rho})$ hereafter.

\subsection{The Sensing-optimal $p_{\RM{X}}(\M{X})$}
Using $\rv{P}_{\rm d}(\rv{\rho})$ as the sensing performance metric, the problem of designing the sensing-optimal $p_{\RM{X}}(\M{X})$ can be formulated as
\begin{equation}\label{detection_optimal}
\max_{p_{\RM{X}}(\M{X})}~~\mathbb{E}\left\{\rv{P}_{\rm d}(\rv{\rho})\right\},~~{\rm s.t.}~~\mathbb{E}\left\{{\rm Tr}(\RM{R}_{\RM{X}})\right\}\leq P.
\end{equation}
Comparing \eqref{detection_optimal} with \eqref{determinisitic_sensing_cost}, we see that $e(\RM{X}) = -\rv{P}_{\rm d}(\rv{\rho})$, the resource $c(\RM{X})$ is the transmit power ${\rm Tr}(\RM{R}_{\RM{X}})$, and $P=C$ is the expected power constraint. According to Proposition \ref{prop:optimal_px}, we may first determine the optimal conditional distribution $p_{\RM{X}|{\rm Tr}(\RM{R}_{\RM{X}})}(\M{X}|{\rm Tr}(\M{R}_{\M{X}}))$. Observe that in the problem \eqref{detection_optimal}, both the objective function and the constraint are related to $\RM{X}$ only via $\RM{R}_{\RM{X}}$, and hence it suffices for us to find $p_{\RM{X}|{\rm Tr}(\RM{R}_{\RM{X}})}(\M{X}|{\rm Tr}(\M{R}_{\M{X}}))$. This can be achieved by considering the deterministic optimization problem
\begin{equation}\label{constrained_optimal_pd}
\max_{\RM{X}}~\rv{P}_{\rm d}(\rv{\rho}),~~{\rm s.t.}~{\rm Tr}(\RM{R}_{\RM{X}}) = P,
\end{equation}
which is equivalent to
\begin{equation}\label{eigenvalue_optimal_pd}
\max_{\RM{X}}~{\rm Tr}(\M{H}_{\rm s}\RM{R}_{\RM{X}}\M{H}_{\rm s}^{\rm H}),~~{\rm s.t.}~{\rm Tr}(\RM{R}_{\RM{X}})= P,
\end{equation}
since $\rv{P}_{\rm d}(\rv{\rho})$ is a monotonically increasing function of $\rv{\rho}$, which is in turn monotonically increasing with respect to ${\rm Tr}(\M{H}_{\rm s}\RM{R}_{\RM{X}}\M{H}_{\rm s}^{\rm H})$. The problem \eqref{eigenvalue_optimal_pd} admits the following optimal solution structure
\begin{equation}\label{optimal_rx_structure}
\RM{R}_{\RM{X}}^{\rm opt} = \M{U}_{\max}\RM{P}\M{U}_{\max}^{\rm H},
\end{equation}
where $\RM{P}\succeq\M{0}$ is a diagonal matrix satisfying ${\rm Tr}(\RM{P})=P$, while $\M{U}_{\max}$ contains the eigenvectors of $\M{H}_{\rm s}^{\rm H}\M{H}_{\rm s}$ corresponding to the largest eigenvalue of $\M{H}_{\rm s}^{\rm H}\M{H}_{\rm s}$ (denoted as $\lambda_{\max}(\M{H}_{\rm s}^{\rm H}\M{H}_{\rm s})$). Thus we have
\begin{equation}
\begin{aligned}
&p_{\RM{R}_{\RM{X}}|{\rm Tr}(\RM{R}_{\RM{X}})}(\M{R}_{\RM{X}}|{\rm Tr}(\M{R}_{\RM{X}})=P) \\
&\hspace{3mm}= \int \delta\left(\M{R}_{\RM{X}}-\M{U}_{\max}\M{P}_P\M{U}_{\max}^{\rm H}\right)p_{\RM{P}_P}(\M{P}_P){\rm d}\M{P}_P,
\end{aligned}
\end{equation}
where $p_{\RM{P}_P}(\M{P}_P)$ can be an arbitrary distribution over positive semidefinite diagonal matrices satisfying ${\rm Tr}(\RM{P}_P)=P$. As an important special case, when the multiplicity of the $\lambda_{\max}(\M{H}_{\rm s}^{\rm H}\M{H}_{\rm s})$ is $1$, we have
\begin{equation}
p_{\RM{R}_{\RM{X}}|{\rm Tr}(\RM{R}_{\RM{X}})}(\M{R}_{\RM{X}}|{\rm Tr}(\M{R}_{\RM{X}})=P) = \delta\left(\M{R}_{\RM{X}}-P\M{u}_{\max}\M{u}_{\max}^{\rm H}\right),
\end{equation}
where $\V{u}_{\max}$ is the unique eigenvector corresponding to $\lambda_{\max}(\M{H}_{\rm s}^{\rm H}\M{H}_{\rm s})$.

Next we will find the sensing-optimal $p_{{\rm Tr}(\RM{R}_{\RM{X}})}({\rm Tr}(\M{R}_{\RM{X}}))$. The scalarization-achievable Pareto front $\widetilde{g}({\rm Tr}(\M{R}_{\RM{X}}))$ can be computed as
\begin{equation}
\widetilde{g}({\rm Tr}(\M{R}_{\RM{X}})) = \left\{
\begin{array}{ll}
P_{\rm d}(\rho), & \hbox{${\rm Tr}(\M{R}_{\RM{X}})\in\widetilde{\Set{G}}$;} \\
{\rm undefined}, & \hbox{otherwise.}
\end{array}
\right.
\end{equation}
The domain $\widetilde{\Set{G}}$ is given by
$$
\begin{aligned}
&\widetilde{\Set{G}} = \Big\{{\rm Tr}(\M{R}_{\RM{X}})\Big|P_{\rm d}(\rho)-\lambda {\rm Tr}(\M{R}_{\RM{X}}) = \\
&\hspace{3mm}\max_{\RM{X}}~\big[P_{\rm d}(\rho)-\lambda {\rm Tr}(\M{R}_{\RM{X}})\big],~\lambda\geq 0\Big\}.
\end{aligned}
$$
According to \eqref{optimal_rx_structure}, we may rewrite $\widetilde{\Set{G}}$ as
\begin{equation}
\widetilde{\Set{G}} = \left\{P\Big|f(P)-\lambda P = \max_P~\left[f(P)-\lambda P\right],~\lambda\geq 0\right\},
\end{equation}
where $f(P)$ is defined as
\begin{equation}
f(P)=\rv{P}_{\rm FA}^{\left(1+\frac{\overline{\rv{A}^2}T}{N_0}\lambda_{\max}(\M{H}_{\rm s}^{\rm H}\M{H}_{\rm s})P \right)^{-1}}.
\end{equation}
Note that $f(P)$ monotonically increases with $P$, but does not have constant concavity. Indeed, we see that $f$ is convex when
\begin{equation}
P < -\frac{\ln \rv{P}_{\rm FA}}{2\alpha^3} - \frac{1}{\alpha} = P_*,
\end{equation}
while being concave when $P>P_*$. In light of this, $\widetilde{\Set{G}}$ may be expressed as
\begin{equation}
\widetilde{\Set{G}} = \left\{P\Big| P=0~{\rm or}~P\geq P_{\rm t}\right\},
\end{equation}
where $P_{\rm t}$ is the value of $P$ satisfying 
\begin{equation}\label{pt_equation}
\frac{{\rm d}f(P)}{{\rm d}P}P+f(0)= f(P).
\end{equation}
Note that $(P_{\rm t},f(P_{\rm t}))$ is the point at which a half-line from the point of origin is tangent to $f(P)$, which is unique. Now the optimal $p_{{\rm Tr}(\RM{R}_{\RM{X}})}({\rm Tr}(\M{R}_{\RM{X}}))$ can be expressed as
\begin{equation}
\begin{aligned}
&p_{{\rm Tr}(\RM{R}_{\RM{X}})}({\rm Tr}(\M{R}_{\RM{X}})) \\
&\hspace{3mm}= \left\{
\begin{array}{ll}
    \frac{P_{\rm t}-P}{P_{\rm t}}\delta\left({\rm Tr}(\M{R}_{\RM{X}})-P_{\rm t}\right) + \frac{P}{P_{\rm t}}\delta\left({\rm Tr}(\M{R}_{\RM{X}})\right), & \hbox{$P \leq P_{\rm t}$;} \\
    \delta\left({\rm Tr}(\M{R}_{\RM{X}})-P_{\rm t}\right), & \hbox{$P\geq P_{\rm t}$.}
\end{array}
\right.
\end{aligned}
\end{equation}
Now we may conclude that the optimal $p_{\RM{R}_{\RM{X}}}(\M{R}_{\RM{X}})$ is given by
\begin{equation}\label{pd_optimal_prx}
\begin{aligned}
&p_{\RM{R}_{\RM{X}}}(\M{R}_{\RM{X}}) \\
&\hspace{3mm}\!=\! \left\{
\begin{array}{ll}
\begin{aligned}
&\frac{P_{\rm t}\!-\!P}{P_{\rm t}}\int \delta\left(\M{R}_{\RM{X}}\!-\!\M{U}_{\max}\M{P}_{P_{\rm t}}\M{U}_{\max}^{\rm H}\right) \\
& p_{\RM{P}_{P_{\rm t}}}(\M{P}_{P_{\rm t}}){\rm d}\M{P}_{P_{\rm t}} + \frac{P}{P_{\rm t}}\delta\left(\M{R}_{\RM{X}}\right),\end{aligned} & \hbox{$P\!\leq\! P_{\rm t}$;}\\
\begin{aligned}&\int \delta\left(\M{R}_{\RM{X}}\!-\!\M{U}_{\max}\M{P}_{P_{\rm t}}\M{U}_{\max}^{\rm H}\right)\\
&\hspace{3mm}p_{\RM{P}_{P_{\rm t}}}(\M{P}_{P_{\rm t}}){\rm d}\M{P}_{P_{\rm t}}\end{aligned}, & \hbox{$P\!\geq\! P_{\rm t}$,}
\end{array}
\right.
\end{aligned}
\end{equation}
for any probability distribution $p_{\RM{P}_P}(\M{P}_P)$ over positive semidefinite diagonal matrices satisfying ${\rm Tr}(\RM{P}_P)=P$. Especially, when the multiplicity of $\lambda_{\max}(\M{H}_{\rm s}^{\rm H}\M{H}_{\rm s})$ is $1$, \eqref{pd_optimal_prx} simplifies to
\begin{equation}\label{pd_optimal_prx_simple}
p_{\RM{R}_{\RM{X}}}(\M{R}_{\RM{X}}) = \left\{
\begin{array}{ll}
\begin{aligned}&\frac{P_{\rm t}-P}{P_{\rm t}}\delta\left(\M{R}_{\RM{X}}-P_{\rm t}\M{u}_{\max}\M{u}_{\max}^{\rm H}\right) \\
&\hspace{3mm}+ \frac{P}{P_{\rm t}}\delta\left(\M{R}_{\RM{X}}\right)\end{aligned}, & \hbox{$P\leq P_{\rm t}$;}\\
\delta\left(\M{R}_{\RM{X}}-P\M{u}_{\max}\M{u}_{\max}^{\rm H}\right), & \hbox{$P\geq P_{\rm t}$.}
\end{array}
\right.
\end{equation}
\begin{remark}[\ac{drt} without convexity/concavity]
From \eqref{pd_optimal_prx_simple} we may observe the \ac{drt} clearly: When $P\geq P_{\rm t}$, the sensing-optimal waveform has a deterministic sample covariance matrix $\RM{R}_{\RM{X}}$, similar to the \ac{crb}-rate tradeoff analyzed in \cite{tit_crb_rate}. When $P\leq P_{\rm t}$, we are effectively working on the non-concave region of $\rv{P}_{\rm d}(\rv{\rho})$, and hence the sensing-optimal $\RM{R}_{\RM{X}}$ can take distinct values. Nevertheless, the \ac{drt} persists since the support of $\RM{R}_{\RM{X}}$ is shrunken from the entire semidefinite cone to two points, namely $\{P_{\rm t}\V{u}_{\max}\V{u}_{\max}^{\rm H},~\M{0}\}$. 
\end{remark}

\begin{figure}[t]
    \centering
    \subfloat[][$\rv{P}_{\rm d}(\rv{\rho})$ and its convex envelope]{
    \centering
    \begin{overpic}[width=.45\columnwidth]{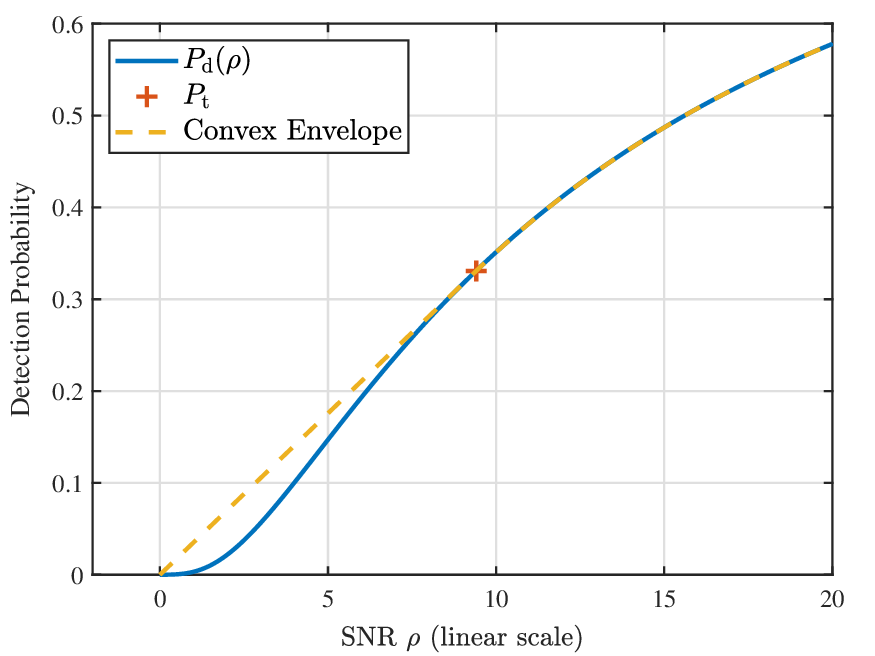}
    \end{overpic}\label{fig:pd_vs_envelope}
    }
    \subfloat[][$p_{\rv{\rho}}(\rho)$ at $P=1<P_{\rm t}$]{
    \centering
    \begin{overpic}[width=.45\columnwidth]{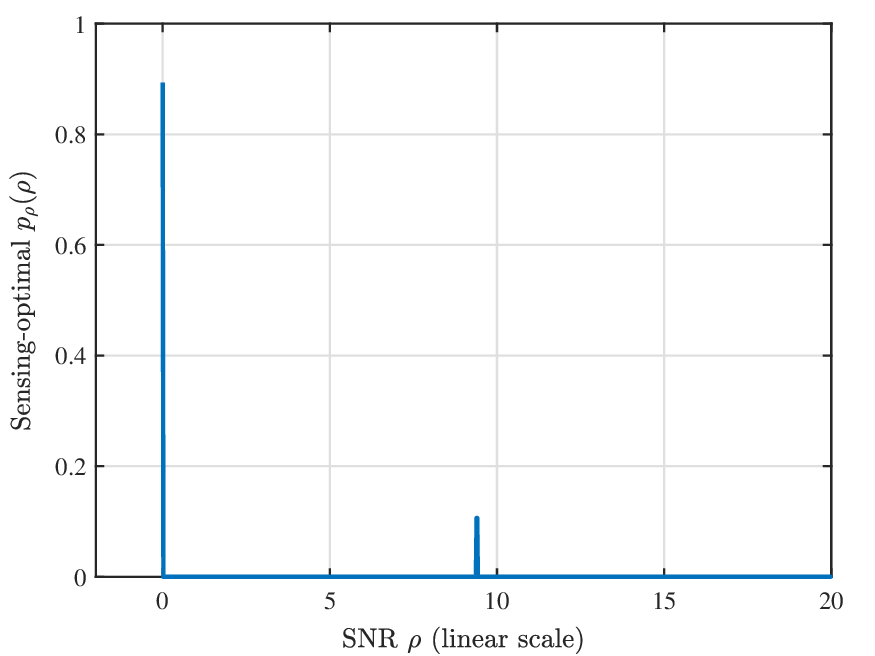}
    \end{overpic}\label{fig:prho_snr_1}
    }\\
    \subfloat[][$p_{\rv{\rho}}(\rho)$ at $P=7<P_{\rm t}$]{
    \centering
    \begin{overpic}[width=.45\columnwidth]{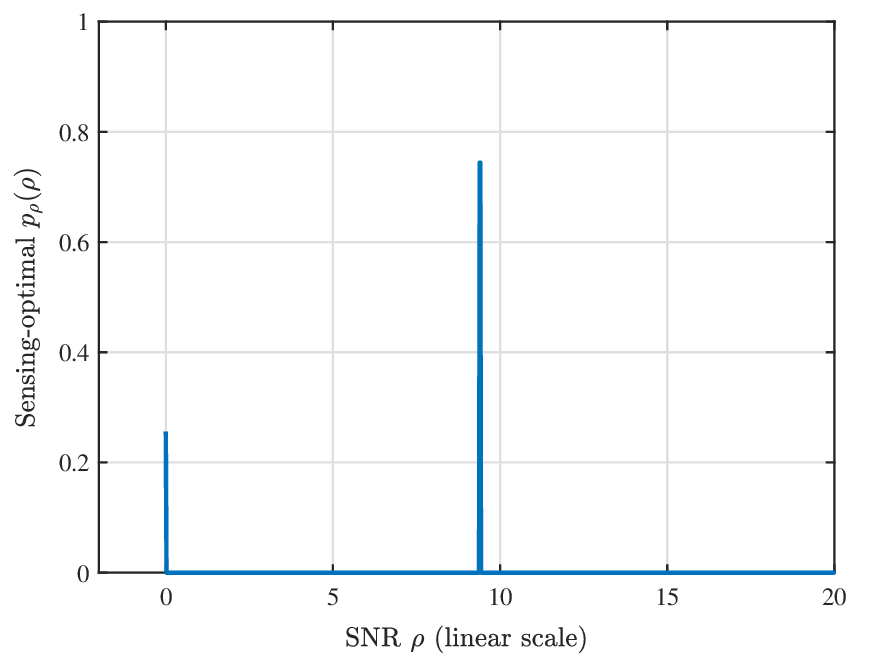}
    \end{overpic}\label{fig:prho_snr_7}
    }
    \subfloat[][$p_{\rv{\rho}}(\rho)$ at $P=15>P_{\rm t}$]{
    \centering
    \begin{overpic}[width=.45\columnwidth]{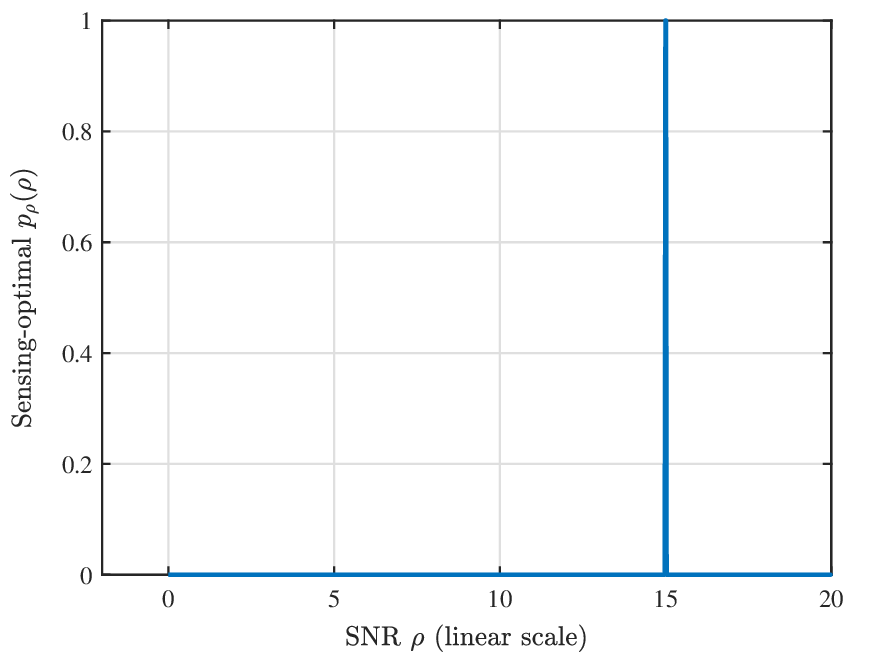}
    \end{overpic}\label{fig:prho_snr_15}
    }
    
    \caption{Numerical results of the detection problem with $\frac{\overline{\rv{A}^2}T}{N_0}\lambda_{\max}(\M{H}_{\rm s}^{\rm H}\M{H}_{\rm s})=1$ and $P_{\rm FA}=10^{-5}$.}
    \label{fig:detection}
\end{figure}

Finally, let us illustrate the previous analytical results using some concrete numerical results. For simplicity, we assume that $\frac{\overline{\rv{A}^2}T}{N_0}\lambda_{\max}(\M{H}_{\rm s}^{\rm H}\M{H}_{\rm s})=1$. Let us further assume that $P_{\rm FA}=10^{-5}$, the corresponding $\rv{P}_{\rm d}(\rv{\rho})$ would then take the form as portrayed in Fig.~\ref{fig:pd_vs_envelope}. In this case, we have $P_{\rm t}\approx 9.4070$ (by solving \eqref{pt_equation} numerically). When we have $P<P_{\rm t}$, the sensing-optimal distribution $p_{\rv{\rho}}(\rho)$ would take the two-point form as in Fig.~\ref{fig:prho_snr_1} and Fig.~\ref{fig:prho_snr_7}. By contrast, when $P\geq P_{\rm t}$, the sensing-optimal distribution is a Dirac delta as shown in Fig.~\ref{fig:prho_snr_15}. These results corroborate \eqref{pd_optimal_prx_simple}.

\section{Conclusions}
In this paper, we have investigated the \ac{drt} between sensing and communication, by analyzing the solution sparsity at the sensing-optimal operating point $P_{\rm SC}$ of the generic capacity-distortion region. Our results suggest that the \ac{drt} originates from the solution sparsity of linear function programming, with the scenarios having convex sensing performance metrics (investigated in previous contributions) being special cases. In particular, we have conducted a case study on the binary target detection task in \ac{mimo} radars, where the sensing performance metric -- detection probability -- is non-convex, for which we have provided explicitly the expressions of the sensing-optimal input waveform distributions. Our results may shed light on the design principles of general practical \ac{isac} systems.

\appendix
\section{Proof of Proposition 1}\label{proof_optimal_px}
\begin{proof}
We first rewrite the optimization problem \eqref{general_sensing_cost} as follows
\begin{subequations}
\begin{align}
\min_{p_{\RM{X}}(\M{X})}&~~\int e(\M{X})p_{\RM{X}}(\M{X}){\rm d}\M{X}\\
{\rm s.t.}&~~\int c(\M{X})p_{\RM{X}}(\M{X}){\rm d}\M{X}\leq C,\\
&~~p_{\RM{X}}(\M{X})\geq 0~\forall \RM{X},~\int p_{\RM{X}}(\M{X}){\rm d}\M{X}=1.
\end{align}
\end{subequations}
The Lagrangian functional reads
$$
\begin{aligned}
&L[p_{\RM{X}}(\M{X}),\lambda_1,\lambda_2,\nu(\M{X})] \\
&\hspace{3mm}= \int e(\M{X})p_{\RM{X}}(\M{X}){\rm d}\M{X}+\lambda_1 \left(\int p_{\RM{X}}(\M{X}){\rm d}\M{X}-1\right) \nonumber\\
&\hspace{5mm}+\lambda_2\left(\int c(\M{X})p_{\RM{X}}(\M{X}){\rm d}\M{X}-C\right) \\
&\hspace{5mm}- \int \nu(\M{X})p_{\RM{X}}(\M{X}){\rm d}\M{X}.
\end{aligned}
$$
The corresponding KKT conditions are as follows
\begin{subequations}\label{KKT_general}
\begin{align}
    &\frac{\partial L}{\partial p_{\RM{X}}(\M{X})} \!=\! e(\M{X})-\nu(\M{X})\!+\!\lambda_1\!+\!\lambda_2c(\M{X}) \!=\! 0~\forall \M{X}, \label{gradient3} \\
    &\nu(\M{X})\geq 0~\forall \M{X},~\lambda_2\geq 0, \label{dual_feasibility3} \\
    &\nu(\M{X})p_{\RM{X}}(\M{X})=0~\forall \M{X},~p_{\RM{X}}(\M{X})\geq 0~\forall \M{X},\\
    &\int p_{\RM{X}}(\M{X}) {\rm d}\M{X} = 1, \label{complementary3}\\
    &\int c(\M{X})p_{\RM{X}}(\M{X}){\rm d}\M{X} \leq C,\\
    &\lambda_2\left(\int c(\M{X})p_{\RM{X}}(\M{X}){\rm d}\M{X}-C\right)=0.
\end{align}
\end{subequations}
From \eqref{gradient3}-\eqref{complementary3} we obtain
\begin{subequations}
\begin{align}
\forall \M{X},~-e(\M{X})&\leq \lambda_1+\lambda_2c(\M{X}), \label{upperbound3}\\
\exists \M{X},~-e(\M{X})&=\lambda_1+\lambda_2 c(\M{X}), \\
{\rm supp}(p_{\RM{X}}(\M{X})) &= \{\M{X}|-e(\M{X})=\lambda_1+\lambda_2c(\M{X})\}. \label{support3}
\end{align}
\end{subequations}
The results \eqref{upperbound3} and \eqref{support3} further imply that
\begin{equation}\label{support_condition1}
\begin{aligned}
&{\rm supp}(p_{\RM{X}}(\M{X}))= \Big\{\M{X}|e(\M{X})+\lambda_2 c(\M{X})=\\
&\hspace{3mm}\min_{\M{X}} \big[e(\M{X})+\lambda_2 c(\M{X})\big] =-\lambda_1\Big\}
\end{aligned}
\end{equation}
holds for some appropriate choice of $\lambda_1$ and $\lambda_2$. This implies that the optimal $p_{\rv{c}(\RM{X})}(c(\M{X}))$ should satisfy
\begin{equation}
\begin{aligned}
&{\rm supp}(p_{\rv{c}(\RM{X})}(c(\M{X}))) = \Big\{c(\M{X})|e(\M{X})+\lambda_2 c(\M{X})=\\
&\hspace{3mm}\min_{\M{X}} \big[e(\M{X})+\lambda_2 c(\M{X})\big]=-\lambda_1\Big\},
\end{aligned}
\end{equation}
and that
\begin{equation}\label{within_g}
{\rm supp}(p_{\RM{X}}(\M{X}))\subseteq \left\{\M{X}|e(\M{X})=\tilde{g}(c(\M{X}))\right\}.
\end{equation}
Since $\tilde{g}(\cdot)$ is constituted by a part of the Pareto front, we have that
\begin{equation}\label{pareto_meaning}
\tilde{g}(\xi_0) =\min_{\xi\in\widetilde{\Set{G}}}~\tilde{g}(\xi),~{\rm s.t.}~\xi\leq \xi_0.
\end{equation}
We observe that when $C\in\widetilde{\Set{G}}$, it follows from \eqref{within_g} and \eqref{pareto_meaning} that
$$
p_{\rv{c}(\RM{X})}(c(\M{X})) = \delta\left(c(\M{X})-C\right).
$$
Another special case is that, when $C \geq \max_{\xi\in\widetilde{\Set{G}}} ~\xi$, since $\tilde{g}(\xi)$ is decreasing with respect to $\xi$, the optimal $p_{\rv{c}(\RM{X})}(c(\M{X}))$ is given by
$$
p_{\rv{c}(\RM{X})}(c(\M{X})) = \delta\left(c(\M{X})-\max_{\xi\in\widetilde{\Set{G}}}~\xi\right).
$$
When $C\notin\widetilde{\Set{G}}$ and $C<\max_{\xi\in\widetilde{\Set{G}}}~\xi$, the optimal $p_{\rv{c}(\RM{X})}(c(\M{X}))$ should satisfy
\begin{subequations}\label{KKT_scalar_trace}
\begin{align}
p_{\rv{c}(\RM{X})}(c(\M{X}))= \mathop{\rm argmin}_{p(\xi)}&~~ \int \tilde{g}(\xi) p(\xi) {\rm d}\xi\\
{\rm s.t.}&~~\int \xi p(\xi){\rm d}\xi \leq C,\\
&~~p(\xi)\geq 0~\forall\xi\geq 0,\\
&~~\int p(\xi){\rm d}\xi=1.
\end{align}
\end{subequations}
Using again the decreasing property of $\tilde{g}(\xi)$, the resource constraint should be active in order to ensure optimality, namely
\begin{equation}\label{active_constraint}
\int \xi p(\xi){\rm d}\xi=C.
\end{equation}
Using \eqref{active_constraint} and the KKT conditions of \eqref{KKT_scalar_trace}, we obtain
\begin{subequations}
\begin{align}
&\lambda\geq 0,~\mu\geq 0, \label{nonnegative}\\
&\forall \xi \geq 0,~-\tilde{g}(\xi)\leq \lambda+\mu\xi, \label{gradient_scalar}\\
&\exists \xi \geq 0,~-\tilde{g}(\xi) = \lambda+\mu\xi, \label{existence_scalar}\\
&\min_{\xi\in\Set{S}[\tilde{g},C]}~\xi \leq P\leq \max_{\xi\in\Set{S}[\tilde{g},C]}~\xi, \label{intersections_scalar}
\end{align}
\end{subequations}
where $\Set{S}[\tilde{g},C]$ is defined as $\Set{S}[\tilde{g},C] = \left\{\xi|-\tilde{g}(\xi)=\lambda+\mu\xi\right\}$. Next we show that the set $\Set{S}[\tilde{g},C]$ is uniquely determined given $\tilde{g}$ and $C$.
\begin{lemma}\label{lem:uniqueness2}
For any $\tilde{g}(\xi)$ being monotonically decreasing (not necessarily strictly), exactly one of the following statements holds:
\begin{enumerate}
\item[1)] $\Set{S}[\tilde{g},C]=\{C\}$, and $\tilde{g}(\xi)$ is not differentiable at $\xi=C$;
\item[2)] There exists one and only one pair of $(\lambda,\mu)$ satisfying the conditions \eqref{nonnegative}-\eqref{intersections_scalar}.
\end{enumerate}
\begin{proof}
First consider the case of $|\Set{S}[\tilde{g},C]|=1$. From \eqref{intersections_scalar} we see that $\Set{S}[\tilde{g},C]=\{C\}$. If $\tilde{g}(\xi)$ is not differentiable at $\xi=C$, statement 1) holds. Otherwise, when $\tilde{g}(\xi)$ is differentiable at $\xi=C$, $\mu$ is uniquely determined by $\mu=\tilde{g}'(C)$, and then $\lambda$ can also uniquely determined by $\tilde{g}(C)=\tilde{g}'(C)C+\lambda$. When $|\Set{S}[\tilde{g},C]|>1$, one may arbitrarily pick two distinct elements $\xi_1\in\Set{S}[\tilde{g},C]$ and $\xi_2\in\Set{S}[\tilde{g},C]$, such that $\lambda$ and $\mu$ are uniquely determined by
$$
-\tilde{g}(\xi_1) = \lambda+\mu \xi_1,~\tilde{g}(\xi_2) = \lambda+\mu \xi_2.
$$
Hence the proof is completed.
\end{proof}
\end{lemma}

Observe that in either case described in Lemma \ref{lem:uniqueness2}, the set $\Set{S}[\tilde{g},C]$ is unique. This is certainly true for Case 1); As for Case 2), $\Set{S}[\tilde{g},C]$ is readily obtained from its definition once $(\lambda,\mu)$ is specified. Thus we may conclude that the optimal $p_{\rv{c}(\RM{X})}(c(\M{X}))$ is indeed determined by the conditions \eqref{optimal_pcx}-\eqref{setg}.

Now that the optimal $p_{\rv{c}(\RM{X})}(c(\M{X}))$ is determined, we proceed to construct the optimal conditional distribution $p_{\RM{X}|\rv{c}(\RM{X})}(\M{X}|c(\M{X})=\xi)$ for every $\xi \in{\rm supp}(p_{\rv{c}(\RM{X})}(c(\M{X})))$. It turns out that this is relatively straightforward, since all optimal solutions to the deterministic optimization problem
\begin{equation}
\min_{\M{X}}~e(\M{X}),~{\rm s.t.}~c(\M{X})=\xi
\end{equation}
produce the same objective function value, and hence can be assigned arbitrary probability densities as long as the normalization condition is satisfied. Upon denoting this optimal solution set as $\Set{S}_{\rm opt}(\xi)$, we may conclude that, given an expected resource budget $C$, the sensing-optimal distribution $p_{\RM{X}}(\M{X})$ should satisfy
\begin{equation}
p_{\RM{X}}(\M{X}) = \int p_{\RM{X}|\rv{c}(\RM{X})}(\M{X}|c(\M{X})) p_{\rv{c}(\RM{X})}(c(\M{X})){\rm d} c(\M{X}),
\end{equation}
and
\begin{equation}
{\rm supp}\left(p_{\RM{X}|\rv{c}(\RM{X})}(\M{X}|c(\M{X})=\xi)\right) = \Set{S}_{\rm opt}(\xi).
\end{equation}
This completes the proof.
\end{proof}

\bibliographystyle{IEEEbib}
\bibliography{gdrt}

\end{document}

%% file: SupportDocuments/YFPreamble.tex
\usepackage{amssymb}
\usepackage[cmex10]{amsmath}
\usepackage{overpic}
\usepackage{algorithm}
\usepackage{algorithmic}
\usepackage{multirow}
\usepackage{array}
\usepackage{graphicx}
\usepackage{bm}
\usepackage{color}
\usepackage{tikz}
\usepackage{epstopdf}
\usepackage{stfloats}
\usepackage{cite}
\usepackage{tikz}
\usepackage{amsmath}
\usepackage{psfrag}
\usepackage{acronym}
\usepackage[caption=false,font=scriptsize]{subfig}
\usepackage{enumerate}
\usetikzlibrary{arrows}
\usetikzlibrary{decorations}

\definecolor{BLUE}{rgb}{0,0,1}

\newtheorem{proposition}{Proposition}
\newtheorem{remark}{Remark}
\newtheorem{lemma}{Lemma}

\acrodef{aoa}[AOA]{angle-of-arrival}
\acrodef{acf}[ACF]{autocorrelation function}
\acrodef{bcrb}[BCRB]{Bayesian Cram\'{e}r-Rao bound}
\acrodef{bp}[BP]{belief propagation}
\acrodef{cdi}[CDI]{cooperative dilution intensity}
\acrodef{cl}[CL]{cooperative localization}
\acrodef{cdf}[CDF]{cumulative distribution function}
\acrodef{crb}[CRB]{Cram\'{e}r-Rao bound}
\acrodef{crlb}[CRLB]{Cram\'{e}r-Rao lower bound}
\acrodef{dof}[DoF]{degree of freedom}
\acrodef{dct}[DCT]{discrete cosine transform}
\acrodef{dpeb}[DPEB]{directional position error bound}
\acrodef{fim}[FIM]{Fisher information matrix}
\acrodef{efim}[EFIM]{equivalent Fisher information matrix}
\acrodef{ici}[ICI]{information coupling intensity}
\acrodef{icrb}[ICRB]{inverse CRB}
\acrodef{iid}[i.i.d.]{independently and identically distributed}
\acrodef{isac}[ISAC]{Integrated Sensing and Communication}
\acrodef{mse}[MSE]{mean-squared error}
\acrodef{pdf}[PDF]{probability density function}
\acrodef{peb}[PEB]{position error bound}
\acrodef{speb}[SPEB]{squared position error bound}
\acrodef{pll}[PLL]{phase-locked loop}
\acrodef{rbs}[RBS]{reference broadcast synchronization}
\acrodef{rhs}[RHS]{right hand side}
\acrodef{rii}[RII]{ranging information intensity}
\acrodef{rss}[RSS]{received signal strength}
\acrodef{rc}[RC]{ranging coefficient}
\acrodef{speb}[SPEB]{squared position error bound}
\acrodef{toa}[TOA]{time-of-arrival}
\acrodef{tdoa}[TDOA]{time-difference-of-arrival}
\acrodef{tpsn}[TPSN]{time synchronization protocol for sensor network}
\acrodef{vmp}[VMP]{variational message passing}
\acrodef{wsn}[WSN]{wireless sensor network}
\acrodef{efim}[EFIM]{equivalent Fisher information matrix}
\acrodef{dio}[DIO]{distance-information-only}
\acrodef{aio}[AIO]{angle-information-only}
\acrodef{saaf}[SAAF]{squared array aperture function}
\acrodef{snc}[S\&C]{sensing and communications}
\acrodef{uoa}[UOA]{uniformly oriented array}
\acrodef{rgg}[RGG]{random geometric graph}
\acrodef{rms}[RMS]{root-mean-square}
\acrodef{snr}[SNR]{signal-to-noise ratio}
\acrodef{eoc}[EoC]{efficiency of cooperation}
\acrodef{npi}[NPI]{nominal position information}
\acrodef{gnss}[GNSS]{global navigation satellite system}
\acrodef{mimo}[MIMO]{multiple-input multiple-output}
\acrodef{mcs}[MCS]{minimally constrained system}
\acrodef{zzb}[ZZB]{Ziv-Zakai bound}
\acrodef{wwb}[WWB]{Weiss-Weinstein lower bound}
\acrodef{nlos}[NLOS]{non-light-of-sight}
\acrodef{mmse}[MMSE]{minimum mean squared error}
\acrodef{uav}[UAV]{unmanned aerial vehicle}
\acrodef{ppp}[PPP]{Poisson point process}
\acrodef{bpp}[BPP]{binomial point process}
\acrodef{cln}[CLN]{cooperative location-aware network}
\acrodef{pdr}[PDR]{pedestrian dead reckoning}
\acrodef{ml}[ML]{maximum likelihood}
\acrodef{map}[MAP]{maximum \textit{a posteriori}}
\acrodef{isac}[ISAC]{integrated sensing and communications}
\acrodef{6g}[6G]{sixth-generation}
\acrodef{itu}[ITU]{international telecommunication union}
\acrodef{snc}[S\&C]{sensing and communication}
\acrodef{crb}[CRB]{Cram\'{e}r-Rao bound}
\acrodef{cfar}[CFAR]{constant false alarm rate}
\acrodef{st}[ST]{subspace tradeoff}
\acrodef{drt}[DRT]{deterministic-random tradeoff}
\acrodef{dof}[DoF]{degree of freedom}

\acrodefplural{dof}[DoFs]{degrees of freedom}